\documentclass[11pt,a4paper]{article}

\usepackage[left=2.8cm,right=2.8cm,top=3cm,bottom=3.5cm,a4paper]{geometry}
\usepackage{latexsym}
\usepackage{amsmath}
\usepackage{amssymb}
\allowdisplaybreaks
\usepackage[dvips]{hyperref} 

\newcommand{\sgn}{\mathrm{sgn}}

\renewcommand{\thefootnote}{\fnsymbol{footnote}}

\begin{document}
\begin{center}
{{{\Large {\bf Central Charges in 2d Reduced Cosmological Massive Gravity}}}}\\[10mm]
{Wontae Kim$^\mathrm{a,b,}$\footnote{E-mail address: wtkim@sogang.ac.kr}, and
Edwin J. Son$^\mathrm{a,}$\footnote{E-mail address: eddy@sogang.ac.kr}}\\[8mm]

{{${}^\mathrm{a}$\it Department of Physics, Sogang University, Seoul, 121-742, Korea\\[0pt]
${}^\mathrm{b}$Center for Quantum Space-time, Sogang University, Seoul, 121-742, Korea}\\[0pt]
}
\end{center}
\vspace{2mm}
\begin{abstract}
In the dimensionally reduced model of the 2+1 dimensional cosmological massive gravity, we obtain the central charges of the two types of the black hole based on the entropy function method.
One is for the BTZ black hole and the other one is actually for the warped AdS$_3$ black hole.
\end{abstract}
\vspace{5mm}



\vspace{1.5cm}

\begin{flushright}
Typeset Using \LaTeX
\end{flushright}
\newpage
\renewcommand{\thefootnote}{\arabic{footnote}}
\setcounter{footnote}{0}

\section{Introduction}
One of the microscopic description of the black hole entropy is to use the celebrated anti-de Sitter (AdS)/conformal field theory (CFT) correspondence~\cite{maldacena}, where the essential ingredients are the near horizon AdS region of the black hole geometry and the low-energy CFT describing the underlying branes.
In particular, there has been much interests in three-dimensional AdS, since a number of black hole solutions have near horizon AdS$_3$ factors and there are powerful results constraining the spectrum of states for its dual theory, CFT$_2$~\cite{kraus}.
For instance, the well-known solution whose asymptotic geometry is AdS$_3$ is the Ba\~nados-Teitelboim-Zanelli (BTZ) black hole~\cite{btz} in which the central charge is calculated as $c=3\ell/2G_3$ with the Brown-Henneaux boundary condition~\cite{bh}, in the Einstein-Hilbert (EH) action with the three-dimensional gravitational constant $G_3$ and a negative cosmological constant $\Lambda=-\ell^{-2}$.

Recently, considering the gravitational Chern-Simons term~\cite{djt} as a correction to the EH action with a negative cosmological constant, which is called the cosmological topologically massive gravity (CTMG), the left moving and right moving central charges for the BTZ black hole which is also a solution to CTMG have been recast as $c_{L,R} = (3\ell/2G_3)(1\mp1/\mu\ell)$, where $\mu$ is the graviton mass~\cite{kl}.
Moreover, for the warped AdS$_3$ black hole as another solution to CTMG, the central charges have been conjectured as $c_L=12\mu\ell^2/G_3(\mu^2\ell^2+27)$ and $c_R=(15\mu^2\ell^2+81)/G_3\mu(\mu^2\ell^2+27)$~\cite{alpss}, where the right moving central charge has been calculated by the Virasoro algebra~\cite{cd}.
Furthermore, Kerr/CFT and extreme black hole/CFT correspondences have recently been under investigation~\cite{ghss,hmns}. Then, the entropies for all of the above cases can be easily given by the Cardy formula~\cite{cardy}.

Apart from the Cardy formula, there is another useful method to get the entropy with the help of the near horizon AdS region. It is called the entropy function method~\cite{sen}, which is based on the Wald entropy formula~\cite{wald}.
It has been shown by using the entropy function that the entropy is proportional to the dilaton value at the horizon in two dimensions even in the presence of higher curvature terms~\cite{hkos}.
Then, the entropy function has been shown to be useful to get the entropy of the CTMG with the dimensional reduction~\cite{ss}, and the 2d reduced CTMG can be used to obtain the left moving central charge of warped AdS$_3$ as was conjectured in Ref.~\cite{alpss} as well as the left and right moving central charges of pure AdS$_3$ case~\cite{afm}.

On the other hand, a new massive gravity (MG) in three dimensions has recently been proposed, including fourth order derivative terms instead of the gravitational Chern-Simons term~\cite{bht}, and the MG admits both the BTZ and warped AdS$_3$ black hole solutions, whose entropies are given by
\begin{equation}
\label{ent:clement}
S_\mathrm{BTZ} = \frac{2\pi r_h}{4G_3} \left( 1 - \frac{1}{2m^2\ell^2} \right),
\quad
S_\mathrm{wAdS_3} = \frac{2\pi r_h}{4G_3} \left( 1 - \frac{m^2\ell^2-6}{5m^2\ell^2} \right),
\end{equation}
respectively, from the Wald entropy formula, where $m$ is a new mass parameter giving a coupling of the higher curvature term and the parameter $\ell$ is now set by $\ell^{-2} = 2m^2 \left( -1 \pm \sqrt{1-\Lambda/m^2} \right)$ for the BTZ and $\ell^{-2} = (2m^2/63) \left( 39 \mp 10 \sqrt{3(5+7\Lambda/m^2)} \right)$ for the warped AdS$_3$ black hole~\cite{clement}.
The Brown-Henneaux boundary condition gives us the central charge for the BTZ black hole as~\cite{ls}
\begin{equation}
c = \frac{3\ell}{2G_3} \left( 1 - \frac{1}{2m^2\ell^2} \right).
\end{equation}
For the warped AdS$_3$ case, however, the central charge has not been obtained as far as we know.
So, we are going to make a conjecture on the central charge for the warped AdS$_3$ black hole in the MG, which will be seen in the form of
\begin{equation}
c = \frac{24\sqrt2 (1+3/2m^2\ell^2)^{3/2}}{\sqrt{5} G_3 m [1+63/2m^2\ell^2]}, 
\end{equation}
where it will be given by the entropy function method and the Cardy formula.

The rest of the present paper is organized as follows.
In Sec.~\ref{sec:KK}, the MG action is reduced to two-dimensional Maxwell-dilaton gravity by the Kaluza-Klein reduction. The resulting action is much more complicated than that of the CTMG case. It can be shown that there exist four possible AdS$_2$ solutions characterized by the constant dilaton, remaining the ambiguity in the sign of the gauge field.
They will be determined explicitly by the entropy function method in Sec.~\ref{sec:S-ftn}. 
Among the four solutions, the two correspond to the first solution (BTZ black hole) and the other two correspond to the second solution (a kind of warped black hole) in Ref.~\cite{clement}.
The entropy for each solution is given by the explicit form, in which the central charge is read from the Cardy formula.
Finally, some comments and discussion will be given in Sec.~\ref{sec:discussion}.

\section{2d reduced MG}
\label{sec:KK}
Instead of the third order of gravitational Chern-Simons term in the CTMG, the MG action contains fourth order derivative terms which is added to the EH action with a cosmological constant $\Lambda$.
It is explicitly given by~\cite{bht}
\begin{align}
& I_{MG} = I_{EH}+I_{K}, \label{act:MG} \\
& I_{EH} = \frac{1}{16\pi G_{3}} \int d^3x \sqrt{-\mathcal{G}} \left[ \mathcal{R}-2\Lambda\right], \\
& I_{K} = -\frac{1}{16\pi G_{3} m^2} \int d^3x \sqrt{-\mathcal{G}} \left[ \mathcal{R}^{MN} \mathcal{R}_{MN} - \frac38 \mathcal{R}^2 \right],
\end{align}
where $G_{3}$ is a three-dimensional Newton constant, $m$ is a mass parameter, and the product, $G_{3} m$, is dimensionless. 
Note that in the Ho\v{r}ava-Lifshitz gravity with $z=4$ in 3+1 dimensions, the MG can be considered as a Euclidean action $W$ in the spatial 3 dimensions~\cite{horava}, where the details will not be given here.
The curly $\mathcal{G}$ and $\mathcal{R}$ represent the metric and curvature in three dimensions in contrasted with the usual $g$ and $R$ in two dimensions.
Note that there are two types of black hole solutions in three dimensions~\cite{clement}: one is the BTZ black hole solution, which is called \emph{the first solution}, and the other is one including warped AdS$_3$ black hole, \emph{the second solution}.

Now, we obtain a two-dimensional Maxwell-dilaton gravity action from the action~\eqref{act:MG} by the Kaluza-Klein reduction with the metric ansatz,
\begin{equation}
\label{met:KK}
ds_{(3)}^2 = g_{\mu\nu} dx^\mu dx^\nu + \ell^2 e^{2\phi} (dy + A_\mu dx^\mu)^2,
\end{equation}
where $\ell$ is a parameter with the dimension of length, which will be fixed later, and $A_\mu$ is a gauge field in two dimensions.
It has been well-known that the EH action is reduced to the conventional Maxwell-dilaton gravity with a field strength $F_{\mu\nu} = \partial_\mu A_\nu - \partial_\nu A_\mu$~\cite{ao},
\begin{equation}
I_{MDG} = \frac{\ell}{8G_3} \int d^2x \sqrt{-g} e^{\phi} \left[ R - 2 \Lambda - \frac{\ell^2}{4} e^{2\phi} F_{\mu\nu} F^{\mu\nu} \right].
\end{equation}
Then, following the Kaluza-Klein reduction~\cite{mh}, we obtain the following 2d reduced MG action,
\begin{align}
& I_{tot} = I_{MDG}+I_{C}, \label{act:tot} \\
& I_{C} = -\frac{\ell}{16 G_3 m^2} \int d^2x \sqrt{-g} e^{\phi} \left[ -\frac14 R^2
  + R \xi + e^{-2\phi} \Phi + \ell^2 e^{-4\phi} \nabla_\mu (e^{3\phi} F^{\mu\rho}) \nabla_\nu (e^{3\phi} F^{\nu}_{~\rho})
  \right. \nonumber \\
& \qquad \qquad \qquad \qquad \qquad \qquad \qquad \left. - \frac{3\ell^2}{4} e^{\phi} F_{\mu\nu} F^{\mu\nu} \Box e^{\phi} + \frac{21\ell^4}{64} e^{4\phi} ( F_{\mu\nu} F^{\mu\nu} )^2 \right],
\end{align}
where $\xi$ and $\Phi$ are defined by $\xi =R/2 + e^{-\phi} \Box e^{\phi} - (5\ell^2/8) e^{2\phi} F_{\mu\nu} F^{\mu\nu}$, $\Phi = 2 \nabla_\mu \nabla_\nu e^{\phi} \nabla^\mu \nabla^\nu e^{\phi} - ( \Box e^{\phi} )^2$, and we used the following two-dimensional identities:
\begin{align}
& R F_{\mu\nu} F^{\mu\nu} = 2 R_{\mu\nu} F^{\mu\alpha} F^{\nu}_{~\alpha}, \qquad R \Box e^{\phi} = 2 R_{\mu\nu} \nabla^\mu \nabla^\nu e^{\phi}, \\
& F_{\mu\nu} F^{\mu\nu} \Box e^{\phi} = 2 F^{\mu\alpha} F^{\nu}_{~\alpha} \nabla_\mu \nabla_\nu e^{\phi}, \quad ( F_{\mu\nu} F^{\mu\nu} )^2 = 2 F_{\mu\nu} F^{\nu\rho} F_{\rho\sigma} F^{\sigma\mu},
\end{align}
which can be easily seen by using the relations, $R_{\mu\nu} = \frac12 g_{\mu\nu} R$ and $F_{\mu\nu} = - \ell^{-2} \epsilon_{\mu\nu} f$, where the two-dimensional antisymmetric tensor $\epsilon_{\mu\nu}$ is defined by $\epsilon_{01} = +\sqrt{-g}$.

After some tedious calculations, we could obtain the equations of motion, but we will not write down them explicitly because it is too lengthy.
Now, assuming a constant dilaton, we can find AdS$_2$ solution satisfying the following reduced equations:
\begin{align}
& 2 \Lambda e^{\phi} + \frac{1}{2\ell^2} e^{3\phi} f^2 - \frac{1}{2m^2} e^{\phi} \left[ \frac14 R^2 + \frac{5}{2\ell^2} R e^{2\phi} f^2 + \frac{63}{16\ell^4} e^{4\phi} f^4 \right] = 0, \label{eq:trace} \\
& e^{\phi} \left\{ R - 2 \Lambda + \frac{3}{2\ell^2} e^{2\phi} f^2  - \frac{1}{2m^2} \left[ \frac14 R^2 + \frac{15}{4\ell^2} R e^{2\phi} f^2 + \frac{105}{16\ell^4} e^{4\phi} f^4 \right] \right\} = 0. \label{eq:dilaton}
\end{align}
Eliminating $R^2$ terms in Eqs.~\eqref{eq:trace} and \eqref{eq:dilaton}, we get
\begin{equation}
R \left( 1 - \frac{5}{8m^2\ell^2} e^{2\phi} f^2 \right) = 4 \Lambda - \frac{1}{\ell^2} e^{2\phi} f^2 + \frac{21}{16m^2\ell^4} e^{4\phi} f^4, \label{eq:curv}
\end{equation}
so that our 2d reduced model may admit AdS$_2$ solutions.
Next, one can eliminate $R$ in Eqs.~\eqref{eq:trace} and \eqref{eq:dilaton} using Eq.~\eqref{eq:curv}, then the following fourth order equation in $e^{2\phi}f^2$ is obtained,
\begin{equation}
\left[ \left( \frac{e^{2\phi} f^2}{8} + m^2 \ell^2 \right)^2 - m^4 \ell^4 \left( 1 - \frac{\Lambda}{m^2} \right) \right] \left[ \left( \frac{21 e^{2\phi} f^2}{4} - 6 m^2 \ell^2 \right)^2 - 3 m^4 \ell^4 \left( 5 + \frac{7\Lambda}{m^2} \right) \right] = 0. \label{eq:field}
\end{equation}
Note that it can be easily solved as
\begin{align}
& e^{2\phi} f^2 = 8 m^2 \ell^2 \left[ -1 \pm \sqrt{1-\Lambda/m^2} \right], && R = -16 m^2 \left[ -1 \pm \sqrt{1-\Lambda/m^2} \right], \label{sol:1st} \\
& e^{2\phi} f^2 = \frac{4 m^2 \ell^2}{21} \left[ 6 \mp \sqrt{3(5+7\Lambda/m^2)} \right], && R = -2 m^2 \left[ 4 \mp \sqrt{3(5+7\Lambda/m^2)} \right]. \label{sol:2nd}
\end{align}
For each solution, both the positive and negative solutions for $f$ are possible, which is related to the handedness of the black hole rotation in three dimensions.
On the other hand, the solutions~\eqref{sol:1st} and \eqref{sol:2nd} are corresponding to \emph{the first solution} and \emph{the second solution}, respectively.
The double signs in the solutions~\eqref{sol:1st} and \eqref{sol:2nd} are in the same ordering in Eqs.~(3.9) and (3.19) in Ref.~\cite{clement}.
The existence of the AdS$_2$ solution requires $m^2 > 0$, $\Lambda < 0$ or $m^2 \le \Lambda < 0$ for the upper sign of the solution~\eqref{sol:1st}, $m^2 < 0$, $\Lambda \ge m^2$ for the lower sign of the solution~\eqref{sol:1st}, $m^2 > 0$, $-5m^2/7 \le \Lambda < m^2/21$ or $\Lambda < m^2 < 0$ for the upper sign of the solution~\eqref{sol:2nd}, and $m^2 > 0$, $\Lambda \ge -5m^2/7$ for the lower sign of the solution~\eqref{sol:2nd}. So, all the solutions~\eqref{sol:1st} and \eqref{sol:2nd} have the AdS$_2$ geometry under the above conditions. 
More details will be given in the following section with the exact form of solutions.

\section{Entropy function and the central charges}
\label{sec:S-ftn}
The entropy function in two dimensional dilaton gravity with $U(1)$ gauge field gives the black hole entropy which is proportional to the overall `effective' coupling constant associated with the dilaton field~\cite{hkos}.
Moreover, it is useful to find the exact AdS$_2$ solutions as well as the entropy of the corresponding solutions.
To obtain the solution and the entropy with $SO(2,1)$ symmetry, the AdS$_2$ solution can be written in the form of
\begin{equation}
ds^2 = v \left( - r^2 dt^2 + \frac{dr^2}{r^2} \right), \qquad e^{\phi} = u, \qquad F_{rt} = \frac{\varepsilon}{\ell^2}, \label{met:ent}
\end{equation}
where $v$, $u$, and $\varepsilon$ are constants to be determined in terms of the charge $q$ and other parameters.
Then, the entropy function is given by
\begin{align}
\mathcal{S} &= 2\pi \left[ q \varepsilon - \mathfrak{L}(u,v,\varepsilon) \right] \notag \\
  &= 2\pi \left\{ q \varepsilon - \frac{\ell}{8G_3} \left[ -2 u - 2 \Lambda v u + \frac{\varepsilon^2 u^3}{2v \ell^2} - \frac{1}{2m^2} \left( \frac{u}{v} - \frac{5 \varepsilon^2 u^3}{2v^2 \ell^2} + \frac{21\varepsilon^4 u^5}{16v^3 \ell^4} \right) \right] \right\},
\end{align}
where $\mathfrak{L}(u,v,\varepsilon)$ is the Lagrangian density for the given field configuration~\eqref{met:ent}.
Next, the extremum equations are given by
\begin{gather}
2 \Lambda u + \frac{\varepsilon^2 u^3}{2v^2 \ell^2} - \frac{1}{2m^2} \left( \frac{u}{v^2} - \frac{5 \varepsilon^2 u^3}{v^3 \ell^2} + \frac{63\varepsilon^4 u^5}{16v^4 \ell^4} \right) = 0, \label{ext:v} \\
2 + 2 \Lambda v - \frac{3\varepsilon^2 u^2}{2v \ell^2} + \frac{1}{2m^2} \left( \frac{1}{v} - \frac{15 \varepsilon^2 u^2}{2v^2 \ell^2} + \frac{105\varepsilon^4 u^4}{16v^3 \ell^4} \right) =0, \label{ext:u} \\
q - \frac{\ell}{8G_3} \left[ \frac{\varepsilon u^3}{v \ell^2} + \frac{1}{2m^2} \left( \frac{5 \varepsilon u^3}{v^2 \ell^2} - \frac{21\varepsilon^3 u^5}{4v^3 \ell^4} \right) \right] = 0, \label{ext:e}
\end{gather}
which determine the constants $u$, $v$, and $\varepsilon$.
Multiplying Eq.~\eqref{ext:v} by $1/u$, we see that Eqs.~\eqref{ext:v} and \eqref{ext:u} are second order of $\varepsilon^2u^2$. From the two equations~\eqref{ext:v} and \eqref{ext:u}, we have $\varepsilon^2u^2$ in terms of $v$:
\begin{equation}
\varepsilon^2 u^2 = 2v\ell^2 \left( \frac{16 \Lambda v^2 + 6v - 1/m^2}{4v - 5/2m^2} \right). \label{sol:e}
\end{equation}
Plugging this into Eq.~\eqref{ext:u}, we have a fourth order equation in $v$, so that there can be four kinds of solutions at most and each one has both positive and negative solutions for $\varepsilon$ as was already seen in Eq.~\eqref{eq:field}.
After some calculations, we can find four sets of consistent solutions:
\begin{align}
& v = \frac{1}{8m^2 (-1+\sqrt{1-\Lambda/m^2})}, && u = \sqrt{\frac{2G_3 |q| \sqrt{2/|1-\sqrt{1-\Lambda/m^2}|}}{|m| ( 2-\sqrt{1-\Lambda/m^2})}}, \notag \\
  & \varepsilon = \frac{\sgn(q)\ell}{4|m|u}
  \sqrt{\frac{2}{|1-\sqrt{1-\Lambda/m^2|}}}, && \text{for (a) } \left\{
  \begin{aligned} 
    & m^2>0, -3m^2 < \Lambda < 0, \Lambda \ne - \frac{11m^2}{25}, \\
    & \text{or } m^2<0, m^2 \le \Lambda < 0,
  \end{aligned} \right. \label{sol:a} \\
& v = \frac{1}{-8m^2 (1+\sqrt{1-\Lambda/m^2})}, && u = \sqrt{\frac{2G_3 |q| \sqrt{2/(1+\sqrt{1-\Lambda/m^2})}}{|m| ( 2+\sqrt{1-\Lambda/m^2})}}, \notag \\
  & \varepsilon = \frac{\sgn(q)\ell}{4|m|u}
  \sqrt{\frac{2}{1+\sqrt{1-\Lambda/m^2}}}, && \text{for (b) } m^2<0, \Lambda \ge m^2, \label{sol:b} \\
& v = \frac{1}{m^2 [4-\sqrt{3(5+7\Lambda/m^2)}]}, && u = \sqrt{\frac{2G_3 |q| \sqrt{6+\sqrt{3(5+7\Lambda/m^2)}}}{m[4-\sqrt{3(5+7\Lambda/m^2)}]\sqrt{1-\Lambda/m^2}}}, \notag \\
  & \varepsilon = \frac{2\sgn(q) \ell
  \sqrt{6-\sqrt{3(5+7\Lambda/m^2)}}}{\sqrt{21}mu
  [4-\sqrt{3(5+7\Lambda/m^2)}]}, && \text{for (c) } -\frac{5m^2}{7} \le \Lambda < \frac{m^2}{21}, \Lambda \ne -\frac{11m^2}{25}, \label{sol:c} \\
& v = \frac{1}{m^2 [4+\sqrt{3(5+7\Lambda/m^2)}]}, && u = \sqrt{\frac{2G_3 |q| \sqrt{6-\sqrt{3(5+7\Lambda/m^2)}}}{m[4+\sqrt{3(5+7\Lambda/m^2)}]\sqrt{1-\Lambda/m^2}}}, \notag \\
  & \varepsilon = \frac{2\sgn(q) \ell
  \sqrt{6+\sqrt{3(5+7\Lambda/m^2)}}}{\sqrt{21}mu
  [4+\sqrt{3(5+7\Lambda/m^2)}]}, && \text{for (d) } -\frac{5m^2}{7} \le \Lambda < m^2, \label{sol:d}
\end{align}
where $\sgn(q)=+1$ for $q>0$ and $-1$ for $q<0$.
The solutions~\eqref{sol:a} and \eqref{sol:b} are corresponding to the upper and lower signs of \emph{the first solution}, and~\eqref{sol:c} and \eqref{sol:d} to the upper and lower signs of \emph{the second   solution} in Ref.~\cite{clement}, respectively.
Note that \emph{the first solution} admits $m^2<0$ case, but \emph{the second solution} does not, even though the solution~\eqref{sol:2nd} admits $\Lambda < m^2 < 0$. We exclude it, because its entropy turns out to be negative.

It is worthy to note that if the effective cosmological constant is defined by $\mathcal{R}=6\Lambda_{eff}$ by uplifting the solutions to three dimensions with the metric ansatz~\eqref{met:KK}, then we have
\begin{align}
& \Lambda_{eff} = 2m^2 \left( 1 \mp \sqrt{1-\Lambda/m^2} \right),
  && \text{for (a) } - \text{ and (b) } +, \label{cos:eff:1st} \\
& \Lambda_{eff} = - \frac{2m^2}{63} \left( 39 \mp 10
  \sqrt{3(5+7\Lambda/m^2)} \right), && \text{for (c) } - \text{ and (d) } +. \label{cos:eff:2nd}
\end{align}
The parameter $\ell$ in the metric ansatz~\eqref{met:KK} is chosen as $\Lambda_{eff}=-\ell^{-2}<0$.
Then, inverting Eqs.~\eqref{cos:eff:1st} and \eqref{cos:eff:2nd}, the following relations are given:
\begin{align}
& \Lambda = -\frac{1}{\ell^2} \left( 1 + \frac{1}{4m^2\ell^2} \right),
&& \text{for } \left\{
  \begin{aligned}
    & \text{(a) } m^2\ell^2 \le -\frac12 \text{ or } m^2\ell^2 > \mathbf{\frac12}, \\
    & \text{(b) } -\frac12 \le m^2\ell^2 < 0,
  \end{aligned} \right. \label{cos:a,b} \\
& \Lambda = \frac{189-468m^2\ell^2+4m^4\ell^4}{400m^2\ell^4}, &&
\text{for } \left\{
  \begin{aligned}
    & \text{(c) } m^2\ell^2 \ge \frac{21}{26}, \\
    & \text{(d) } 0 < m^2\ell^2 \le \frac{21}{26}.
  \end{aligned} \right. \label{cos:c,d}
\end{align}
Note that $\Lambda$ vanishes for $m^2\ell^2=-1/4$, as seen from Eq.~\eqref{cos:a,b}, which shows that the MG for $m^2<0$ admits the BTZ solution even for $\Lambda=0$.
So, even in the absence of a cosmological constant, the MG admits both the BTZ solution and the warped AdS$_3$ black hole solution, according to the sign of $m^2$.
Actually, no consistent solution is possible with $\Lambda=0$ for the case of (a), but it does not mean that the BTZ solution is excluded because we still have a negative effective cosmological constant for the case of (b).
More discussions for $\Lambda=0$, such as gravitational waves on AdS$_3$, can be found in Ref.~\cite{agh}.

Now, the entropy is given by the entropy function evaluated for each set of extremum solutions:
\begin{align}
& S = 4\pi u \left[ \frac{\ell}{8G_3} \left( 1 - \frac{1}{2m^2\ell^2}
  \right) \right] = 2\pi \sqrt{\frac{|q|\ell^2}{6}}
\sqrt{\frac{3\ell}{2G_3} \left( 1 - \frac{1}{2m^2\ell^2} \right)}, &&
  \text{for (a) and (b)}, \\
& S = 4\pi u \left[ \frac{\ell}{8G_3} \left( 1 - \frac{m^2\ell^2-6}{5m^2\ell^2} \right) \right]
  = 2\pi \sqrt{\frac{|q|\ell^2}{6}} \sqrt{\frac{24\sqrt2
  (1+3/2m^2\ell^2)^{3/2}}{\sqrt{5} G_3 m [1+63/2m^2\ell^2]}}, &&
  \text{for (c) and (d)}.
\end{align}
The entropy is proportional to the dilaton field, that is, the effect of the higher curvature correction terms appear only in the numerical factor, as discussed in Ref.~\cite{hkos}.
Note that the entropy satisfies the two-dimensional area law $S=2u$ with $2\pi\ell/8G_3=1$ for (a,b) $m^2\ell^2\to\pm\infty$ and (c,d) $m^2\ell^2\to6$.
Then, from the Cardy formula $S=2\pi\sqrt{cL_0/6}$ with identifying $L_0=|q|\ell^2$, following Ref.~\cite{afm}, the central charges can be
obtained as
\begin{align}
c &= \frac{3\ell}{2G_3} \left( 1 - \frac{1}{2m^2\ell^2} \right), &&
\text{for (a) and (b)}, \label{c:BTZ} \\
c &= \frac{24\sqrt2 (1+3/2m^2\ell^2)^{3/2}}{\sqrt{5} G_3 m
  [1+63/2m^2\ell^2]}, && \text{for (c) and (d)}. \label{c:nonBTZ}
\end{align}
It is interesting to note that the central charges do not depend on the sign of $q$ which is related to the helicity, so that we have $c_L=c_R=c$ for all cases.
Actually, the helicity-dependence seen in the CTMG disappears in the MG; it was a characteristic of the gravitational Chern-Simons or odd-order derivative terms.
Note that the central charge~\eqref{c:BTZ} is exactly the same with the central charge of the MG with the AdS$_3$ background~\cite{ls}.
Moreover, among \emph{the second solution}, only the upper sign, i.e. solution~\eqref{sol:c}, corresponds to the warped AdS$_3$ black hole, as shown in Ref.~\cite{clement}. So, following the notation in Ref.~\cite{alpss},
\begin{equation}
\Lambda = \frac{9-48\nu^2+4\nu^4}{2\ell^2(-3+20\nu^2)}, \qquad m^2 = \frac{-3+20\nu^2}{2\ell^2},
\end{equation}
the central charge~\eqref{c:nonBTZ} can be written as
\begin{equation}
c = \frac{96\ell\nu^3}{G_3(-3+20\nu^2)(3+\nu^2)}, \qquad \text{for (c)}, \label{c:WAdS}
\end{equation}
where it is different from that of the warped AdS$_3$ black hole in the CTMG.

\section{Discussion}
\label{sec:discussion}
We have reduced the cosmological MG to a two-dimensional dilaton gravity with a non-minimally coupled $U(1)$ gauge field, which admits two types of AdS$_2$ solution: One is corresponding to the BTZ black hole, asymptotically AdS$_3$, and the other one is corresponding to an asymptotically non-AdS$_3$ solution, which includes the warped AdS$_3$ black hole solution, by uplifting to three dimensions.
Then, the central charge for the corresponding solution is obtained by using the Cardy formula.
The BTZ black hole has the same central charge as in the literatures, but the central charge for \emph{the second solution} is unknown.
So, we hope the central charge~\eqref{c:nonBTZ} for \emph{the second solution} will be derived in terms of another direct method, for instance, Brown-Henneaux canonical formulation.

As for the cosmological constant $\Lambda$, it has been shown that without it, the MG still admits both the BTZ solution and the warped AdS$_3$ black hole, according to the sign of the coupling of the higher curvature term $I_K$, and the corresponding entropy~\eqref{ent:clement} satisfies the area law up to a numerical factor.
This shows that the higher curvature term in the MG can play the role of a negative cosmological constant as well as the role of the entropy correction.
This behavior has not been seen in the CTMG.



\vspace{1cm}

{\bf Acknowledgments}

We would like to thank M.~Yoon for the helpful discussion at the early stage of this research and S.~Detournay for the comments on the correct sign of the right moving central charge for warped AdS$_3$ black hole in CTMG.
This work was supported by the Korea Science and Engineering Foundation (KOSEF) grant funded
by the Korea government(MEST) through the Center for Quantum Spacetime(CQUeST) of Sogang
University with grant number R11 - 2005 - 021.


\end{document}